\documentclass[twocolumn,iop]{emulateapj}
\usepackage{apjfonts}
\usepackage{color}

\usepackage{amsmath,graphicx,longtable,hyperref}
\usepackage{natbib,threeparttable,rotating}
\usepackage{ulem}

\newcommand{\etal}{et al.}
\newcommand{\hbeta}{H{$\beta$}}
\newcommand{\halpha}{H{$\alpha$}}
\newcommand{\lya}{Ly\,$\alpha$}
\newcommand{\CIV}{C\,{\sevenrm IV}}

\newcommand{\SiIV}{Si\,{\sevenrm IV}}
\newcommand{\CIII}{C\,{\sevenrm III]}}

\newcommand{\AlIII}{Al\,{\sevenrm III}}

\newcommand{\SiIII}{Si\,{\sevenrm III]}}

\def\FeII{Fe\,{\sc ii}}
\def\MgII{Mg\,{\sc ii}}
\def\HeII{He\,{\sc ii}}
\def\CaII{Ca\,{\sc ii}}
\def\CaIIwave{Ca\,{\sc ii}\,$\lambda$3934} 
\def\CaIIb{Ca\,{\sc ii}\,$\lambda$3968}

\newcommand{\OII}{[O{\sevenrm\,II}]}

\newcommand{\NeV}{[Ne\,{\sevenrm\,V}]}
\def \OIII {[O\,{\sc iii}]}

\newcommand{\OIIIab}{[O{\sevenrm\,III}]\,$\lambda\lambda$4959,5007}  
\newcommand{\NII}{[N\,{\sevenrm\,II}]}

\def \NV {N\,{\sc v}}

\newcommand{\SII}{[S{\sevenrm\,II}]}

\newcommand{\OIV}{O\,{\sevenrm IV]}}

\newcommand{\bracket}[1]{\left\langle#1\right\rangle}
   \font\sevenrm=cmr7 scaled 1000
\newcommand{\comments}[1]{}

\def\kms{{\rm km\,s^{-1}}}

\begin{document}

\title{The Sloan Digital Sky Survey Reverberation Mapping Project: Velocity Shifts of Quasar Emission Lines}

\author{Yue Shen$^{1,2,*}$, W.~N. Brandt$^{3,4,5}$, Gordon T. Richards$^{6}$, Kelly D.~Denney$^{7,8,9}$, Jenny E.~Greene$^{10}$, C.~J. Grier$^{3,4}$, Luis C.~Ho$^{11,12}$, Bradley M.~Peterson$^{7,8}$, Patrick Petitjean$^{13}$, Donald P.~Schneider$^{3,4}$, Charling Tao$^{14,15}$, Jonathan R.~Trump$^{3,4,16}$} 

\altaffiltext{1}{Department of Astronomy, University of Illinois at Urbana-Champaign, Urbana, IL 61801, USA; shenyue@illinois.edu}
\altaffiltext{2}{National Center for Supercomputing Applications, University of Illinois at Urbana-Champaign, Urbana, IL 61801, USA}
\altaffiltext{*}{Alfred P. Sloan Research Fellow}
\altaffiltext{3}{Department of Astronomy \& Astrophysics, The Pennsylvania State University, University Park, PA, 16802, USA}
\altaffiltext{4}{Institute for Gravitation and the Cosmos, The Pennsylvania State University, University Park, PA 16802, USA}
\altaffiltext{5}{Department of Physics, 104 Davey Lab, The Pennsylvania State University, University Park, PA 16802, USA }
\altaffiltext{6}{Department of Physics, Drexel University, 3141 Chestnut Street, Philadelphia, PA 19104, USA}
\altaffiltext{7}{Department of Astronomy, The Ohio State University, 140 West 18th Avenue, Columbus, OH 43210, USA}
\altaffiltext{8}{Center for Cosmology and AstroParticle Physics, The Ohio State University, 191 West Woodruff Avenue, Columbus, OH 43210, USA}
\altaffiltext{9}{NSF Astronomy \& Astrophysics Postdoctoral Fellow}
\altaffiltext{10}{Department of Astrophysical Sciences, Princeton University, Princeton, NJ 08544, USA}
\altaffiltext{11}{Kavli Institute for Astronomy and Astrophysics, Peking University, Beijing 100871, China}
\altaffiltext{12}{Department of Astronomy, School of Physics, Peking University, Beijing 100871, China}
\altaffiltext{13}{Institut d'Astrophysique de Paris, Universit\'e Paris 6 et CNRS, 98bis Boulevard Arago, 75014 Paris, France}
\altaffiltext{14}{Aix Marseille Universit\'e, CNRS/IN2P3, CPPM UMR 7346, 13288 Marseille, France}
\altaffiltext{15}{Tsinghua Center for Astrophysics, Tsinghua University, Beijing 100084, China}
\altaffiltext{16}{Hubble Fellow}

\shorttitle{SDSS-RM: Quasar Line Shifts}
\shortauthors{Shen \etal}

\begin{abstract}

Quasar emission lines are often shifted from the systemic velocity due to various dynamical and radiative processes in the line-emitting region. The level of these velocity shifts depends both on the line species and on quasar properties. We study velocity shifts for the line {\it peaks} {(not the centroids)} of various narrow and broad quasar emission lines relative to systemic using a sample of 849 quasars from the Sloan Digital Sky Survey Reverberation Mapping (SDSS-RM) project. The coadded (from 32 epochs) spectra of individual quasars have sufficient signal-to-noise ratio (SNR) to measure stellar absorption lines to provide reliable systemic velocity estimates, as well as weak narrow emission lines. The large dynamic range in quasar luminosity ($\sim 2$\,dex) of the sample allowed us to explore potential luminosity dependence of the velocity shifts. We derive average line peak velocity shifts as a function of quasar luminosity for different lines, and quantify their intrinsic scatter. We further quantify how well the peak velocity can be measured as a function of continuum SNR, and demonstrate there is no systematic bias in the velocity measurements when SNR is degraded to as low as $\sim 3$ per SDSS pixel ($\sim 69\,{\rm km\,s^{-1}}$). Based on the observed line shifts, we provide empirical guidelines on redshift estimation from \OII\,$\lambda3727$, \OIII\,$\lambda5007$, \NeV\,$\lambda3426$, \MgII, \CIII, \HeII\,$\lambda1640$, broad \hbeta, \CIV, and \SiIV, which are calibrated to provide unbiased systemic redshifts in the mean, but with increasing intrinsic uncertainties of 46, 56, 119, 205, 233, 242, 400, 415, and 477 $\kms$, in addition to the measurement uncertainties. These results demonstrate the infeasibility of measuring quasar redshifts to better than $\sim 200\,{\rm km\,s^{-1}}$ with only broad lines. 
\end{abstract}


\section{Introduction}\label{sec:intro}

Extragalactic redshift surveys are increasingly utilizing quasars to probe the large-scale structure of the Universe \citep[e.g.,][]{Dawson_etal_2013}. As such, accurate redshift determination of quasars is important for measuring the correlation functions of quasars and extracting information for cosmology and for quasar distributions inside dark matter halos. In addition, knowing the systemic redshift of the quasar is crucial to studying dynamical processes within its nuclear region and host galaxy, such as outflows.  

There are two common approaches to measuring quasar redshifts from large spectroscopic surveys. The first is a template fitting approach, where the spectra of quasars are cross-correlated with a template spectrum of quasars generated from a training set and a redshift is determined from the best match between the input spectrum and the template \citep[e.g.,][]{Bolton_etal_2012} . A major limitation of this method is that the template is, by definition, an average representation of the quasar population, and does not account for spectral variations in individual quasars, which in turn will affect the redshift accuracy in individual objects. The nominal measurement uncertainties in the redshifts from cross-correlation are often a significant underestimation of the true uncertainties. 

The second approach is to measure the observed wavelengths of individual emission lines in the quasar spectrum to derive line-based redshift estimates \citep[e.g.,][]{Paris_etal_2012}. The advantage of this method is that it does not rely on any template and hence avoids potential systematic biases in the template-based redshifts, if, for example, the template is ill-chosen for the set of quasars under consideration. However, just as for the template-based method, this second approach faces problems due to the diverse kinematics of various emission lines in a quasar spectrum. Therefore one needs to take into account the velocity shifts of individual emission lines with respect to the systemic redshift, at least in an effort to remove any average trends with quasar parameters such as luminosity \citep[e.g.,][]{Hewett_Wild_2010}.  

It has been known for more than 30 years that quasar emission lines are often shifted from the systemic velocity. Most notable is the average blueshift of the high-ionization broad \CIV\ line \citep[e.g.,][]{Gaskell_1982, Wilkes_Carswell_1982,Tytler_Fan_1992,Richards_etal_2002} by hundreds of $\kms$ from systemic, with a strong dependence on luminosity \citep[e.g.,][]{Richards_etal_2011}. Low-ionization broad lines (such as \MgII), however, are observed to have an average velocity closer to the systemic velocity \citep[e.g.,][]{Hewett_Wild_2010}. While high-ionization narrow emission lines such as \OIII\ are not as blueshifted as high-ionization broad lines, they can have line peak shifts\footnote{\OIII\ often shows a blue asymmetry (e.g., the ``blue wing'') that depends on the properties of the quasar \citep[e.g.,][]{Heckman_etal_1981,Peterson_etal_1981,Whittle_1985,Veilleux_1991,Zhang_etal_2011, Shen_Ho_2014}. The velocity offset of this wing \OIII\ component appears to increase with luminosity and Eddington ratio \citep[e.g.,][]{Zhang_etal_2011,Shen_Ho_2014}, and can reach extreme values of $\gtrsim 1000\,{\rm km\,s^{-1}}$ at the highest quasar luminosities \citep[e.g.,][]{Shen_2015,Zakamska_etal_2015}. However, the velocity offset of the core \OIII\ component or the \OIII\ peak is generally more stable and lies within $\sim 50\,\kms$ of the systemic velocity, with only mild dependences on quasar luminosity and Eddington ratio \citep[e.g., fig.\ 2 and fig.\ E2 in][also see Hewett \& Wild 2010]{Shen_Ho_2014}.} of tens of $\kms$, with some extreme cases exceeding a hundred $\kms$ \citep[e.g., the so-called ``blue outliers'', ][]{Zamanov_etal_2002,Boroson_2005,Komossa_etal_2008,Marziani_etal_2015}. These average trends have been generally confirmed with recent large spectroscopic quasar samples \citep[e.g.,][]{Vandenberk_etal_2001, Richards_etal_2002, Shen_etal_2007,Shen_etal_2008, Hewett_Wild_2010, Richards_etal_2011, Shen_etal_2011,Zhang_etal_2011,Shen_Ho_2014}, although differences in measuring the line center and the reference line used for systemic velocity will affect the exact value of the inferred velocity shifts. 

In this work we study the velocity shifts in both broad and narrow emission lines in quasar spectra using the sample of $\sim 850$ quasars from the Sloan Digital Sky Survey Reverberation Mapping project \citep[SDSS-RM,][]{Shen_etal_2015a}. There are two distinctive advantages of the SDSS-RM sample for such a study compared with earlier samples: 1) our sample has very high signal-to-noise ratio per object (see below), allowing measurements of stellar absorption lines (hence reliable systemic velocity measurements) in the low-$z$ subset, as well as measurements of weak emission lines in {\it individual} quasar spectra; 2) our sample covers a larger dynamic range in luminosity than previous SDSS samples, as well as a broad redshift range, enabling an investigation of the velocity shifts as a function of luminosity and the construction of a redshift ladder from low to high redshifts based on lines that are suitable for redshift estimation. 

One purpose of quantifying velocity shifts using the statistical SDSS-RM quasar sample is to provide empirical guidelines on measuring quasar redshifts based on emission lines. We will also quantify the intrinsic scatter in redshift measurements using individual emission lines, which serves as a more realistic estimate of redshift uncertainties for spectroscopic quasar surveys. {Our study shares some similarities with the work by \citet{Hewett_Wild_2010} in motivation and methodology, but with complementary samples. Hewett \& Wild used a large number of SDSS quasars to inter-calibrate different line velocities at different redshifts. Our SDSS-RM sample has a much smaller sample size, but the high SNR spectra allowed us to extend the line velocity measurements to higher redshift and therefore cover a larger dynamic range in quasar luminosity. For example, \citet{Hewett_Wild_2010} were only able to measure the \CaII\ K line reliably at $z<0.4$ with single-epoch SDSS quasar spectra, whereas we can measure \CaII\ K up to $z\gtrsim 1$ for most objects in our sample. }

The paper is organized as follows. In \S\ref{sec:data} we describe the sample used, and in \S\ref{sec:spec} we describe the spectral measurements of velocity shifts among different lines. We present the main results on line shifts in \S\ref{sec:results}, and our empirical recipes for quasar redshift estimation in \S\ref{sec:red}. We summarize our findings in \S\ref{sec:sum}. Throughout the paper we adopt a flat $\Lambda$CDM cosmology with $\Omega_\Lambda=0.7$ and $h_{100}=0.7$ in calculating luminosities. {We use vacuum wavelengths for all spectral analyses and velocity calculations; the SDSS spectra are also stored in vacuum wavelength. However, we use traditional (air wavelength) names to refer to lines at $\lambda>2000$\,\AA\ (such as \OII\ and \OIII). }

\section{Data}\label{sec:data}

The SDSS-RM quasar sample includes 849 broad-line quasars at $0.1<z<4.5$ with a flux limit of $i_{\rm psf}=21.7$. The quasars were selected using a variety of methods combining optical, infrared and variability selection. The details of the target selection are presented in the sample characterization paper (Shen et al., 2016, in preparation). Given the hybrid target selection, the SDSS-RM quasar sample is more representative of the general quasar population than other optical quasar surveys, and samples the diversity of quasars in terms of continuum and emission-line properties (Shen et~al. 2016, in prep.). 

The spectroscopic data used in this work are from the 32 epochs taken in 2014 as part of the SDSS-RM project within the SDSS-III \citep{Eisenstein_etal_2011} Baryon Oscillation Spectroscopic Survey \citep[BOSS,][]{Dawson_etal_2013}, using the BOSS spectrograph \citep{Smee_etal_2013} on the 2.5\,m SDSS telescope \citep{Gunn_etal_2006}. The wavelength coverage of BOSS spectroscopy is $\sim 3650-10,400~\textrm{\AA}$, with a spectral resolution of $R\sim 2000$. Each epoch has a typical exposure time of 2 hrs and the total coadded exposure time is $\sim 60$ hrs. The epoch-by-epoch spectra were pipeline-processed as part of the SDSS-III Data Release 12 \citep{dr12}, followed by a custom flux calibration scheme and improved sky subtraction as described in \citet{Shen_etal_2015a}. The improved spectrophotometry has a nominal absolute accuracy of $\sim 5\%$. All epochs were coadded using the SDSS-III spectroscopic pipeline {\tt idlspec2d}, to produce high signal-to-noise ratio (SNR) spectra for all 849 quasars in the SDSS-RM sample. The median SNR per 69$\,\kms$ pixel across the spectrum is $\sim 30$ in the sample median, sufficient to detect weak stellar absorption line features such as \CaII\ in the low-$z$ subset of the sample \citep[e.g.,][]{Shen_etal_2015b,Matsuoka_etal_2015}. These coadded spectra form the basis of our spectral analysis. 

\section{Spectral Measurements}\label{sec:spec}

\begin{table}
\caption{Line Fitting Parameters}\label{tab:linefit}
\centering
\scalebox{1.0}{
\begin{tabular}{ccccc}
\hline\hline
Line Name & Vacuum Rest Wavelength & $n_{\rm gauss}$ & Complex  & $N_{\rm obj}$\\
& [\AA] & & \\
(1) & (2) & (3) & (4) & (5) \\
\hline
\OIII\,5007  & 5008.24 & 2  & \hbeta & 208  \\
\hbeta$_{\rm br}$ & 4862.68 & 3 & \hbeta & 224 \\
\CaII\,K & 3934.78 & 2 & ... & 446 \\
\OII\,3727 & 3728.48  & 1 & ... & 516 \\
\NeV & 3426.84 & 2 & ... & 627  \\
\MgII & 2798.75 & 3 & ... & 749  \\
\CIII & 1908.73 & 2 & \CIII & 705  \\
\SiIII & 1892.03 & 1 & \CIII   & 688 \\
\AlIII & 1857.40 & 1 & \CIII  & 689 \\
\HeII & 1640.42 & 2  & \CIV & 600  \\
\CIV & 1549.06 & 2  & \CIV & 552  \\
\SiIV & 1399.41$^{*}$ & 2 & \SiIV/\OIV & 454 \\
\hline
\hline\\
\end{tabular}
}
\begin{tablenotes}
      \small
      \item NOTE. --- Fitting parameters for the lines considered in this work. The third column lists the total number of Gaussians used for each line. Multiple lines in the same line complex as specified by name are fit simultaneously. The rest wavelengths of the lines are taken from \citet{Vandenberk_etal_2001}. The last column lists the total number of objects with coverage of the line. 

      $^*$The wavelength of the \SiIV\ line is taken as the arithmetic mean of the central wavelengths of \SiIV\ and \OIV\ (1396.76\,\AA\ and 1402.06\,\AA), but all the results for \SiIV\ can easily be scaled to other definitions of the effective wavelength (e.g., oscillator-strength-weighted average).
\end{tablenotes}
\end{table}

We fit the spectra with parameterized functional models and measure the continuum and line properties from the model fits. This technique has become one of the standard approaches in measuring spectral properties of quasars \citep[e.g.,][]{Shen_etal_2008,Shen_etal_2011,Zhang_etal_2011,Zhang_etal_2013,Stern_Laor_2012a,Stern_Laor_2012b}. Compared with direct measurements from the spectral pixels, this method is more robust against noise and artifacts from the reduction process, although occasionally visual inspection is required to confirm that the model accurately reproduces the data. The details of the spectral fitting procedure with demonstration examples and all spectral measurements for the SDSS-RM sample will be presented in a forthcoming paper (Shen et~al. 2016, in prep); and below we briefly summarize the fitting procedure. 

We first fit a global continuum model to several windows across the rest-frame UV-to-optical that are free of prominent broad or narrow lines. The continuum model consists of a power law, a low-order polynomial, and \FeII\ complexes in both the UV and optical regimes. The low-order polynomial component accounts for the curvature in the continuum shape due to heavy dust reddening in some objects. The \FeII\ complexes are modeled using templates from \citet[][]{Boroson_Green_1992} in the optical and \citet[][]{Vestergaard_Wilkes_2001} in the UV.

The best-fit global continuum (including \FeII) model is subtracted from the full spectrum to produce a line spectrum. We then fit multiple-Gaussian models to the emission lines following our earlier work \citep[e.g.,][]{Shen_etal_2011}. The broad lines are fit by multiple Gaussians and the narrow lines are typically fit with a single Gaussian. However, high-ionization narrow emission lines such as \OIII\ and \NeV\ often show significant blueshifted wings \citep[e.g.,][]{Heckman_etal_1981,Whittle_1985,Zhang_etal_2011,Shen_Ho_2014}; and therefore we use two Gaussians to fit these narrow lines. Table \ref{tab:linefit} lists the fitting parameters for the lines considered in this work, where the rest-frame vacuum wavelength of each line is taken from \citet{Vandenberk_etal_2001}. 

During the continuum and emission-line fits, we remedy for the adverse effects of quasar narrow and broad absorption lines by iterative rejections of pixels that fall below the model fit by 3$\sigma$. We found that this approach generally works well for narrow absorption lines, and also improves the fits for objects with mild broad absorption troughs. But for the few quasars in our sample with strong broad absorption troughs the fits are usually unreliable in the regions affected by the absorption (although the peak can sometimes still be well constrained). These rare cases, while included in our sample, do not affect any of our statistical analyses below. 

This global fitting approach fits the continuum and most of the narrow and broad emission-line regions well. However, for relatively isolated narrow emission lines such as \OII\ and \NeV, the global continuum often overestimates the local continuum level underneath the line. Therefore we use a local continuum fit around these narrow lines to reproduce the correct line profile and continuum level. In addition, we fit the strong stellar absorption line \CaIIwave\ (K) with a double-Gaussian (negative Gaussians to fit the absorption) that well reproduces the absorption. The \CaIIb\ line (H) is not used due to contamination from H$\epsilon$ emission (\CaII\ H line is also weaker than \CaII\ K). \CaII\ absorption is clearly detected in most of the low-$z$ quasars in our high SNR SDSS-RM sample, and serves as a reliable indicator for the systemic velocity.\footnote{Although \CaII\ can be affected by ISM absorption, it should not affect the redshift estimation as the ISM absorption is expected to be at the systemic velocity of the quasar. {In addition, arguably \CaII\ is a better systemic velocity indicator than the often-used \OII\ line, as the latter, while generally shown to yield consistent systemic velocity \citep[e.g.,][]{Kobulnicky_Gebhardt_2000,Weiner_etal_2005},  can potentially suffer from gaseous processes in the quasar narrow-line region \citep[e.g.,][]{Zakamska_Greene_2014,Comerford_Greene_2014}. However, in typical quasar spectra \OII\ is easier to measure than \CaII, since high SNR is required to measure the stellar absorption lines. }} 

Table \ref{tab:linefit} lists the lines considered in this work: these lines were chosen to be the most prominent lines across the rest-frame UV-to-optical part of quasar spectra, and relatively unblended with other lines such that a line center can be unambiguously determined. The \lya\ complex (including \lya\ and \NV) is not considered due to the heavy \lya\ absorption. We also exclude the \halpha\ complex (including \halpha, \NII\ and \SII) from our analysis, as there are only tens of objects with \halpha\ coverage in our sample. Other emission lines are also not considered either due to the difficulty of measuring them cleanly (i.e., in a line complex) or due to the existence of a nearby line better suited for redshift estimation. \\

{Below are some additional notes on individual lines:}

\begin{enumerate}

\item[$\bullet$] The 1400\,\AA\ feature is fit with a double-Gaussian to account for \SiIV\ and \OIV. Since it is difficult to disentangle the many blended lines in this feature, we use the mean wavelength of the two lines listed in table 2 of \citet{Vandenberk_etal_2001} as \SiIV\ and \OIV\ to calculate the velocity shift. {We further treat \SiIV\ and \OIV\ (at 1396.76\,\AA\ and 1402.06\,\AA) as a single broad line, and use ``\SiIV'' to refer to this line complex for simplicity. }

\item[$\bullet$] \HeII\,$\lambda$1640 is fit simultaneously with \CIV\ in the \CIV\ complex, with independent constraints on velocity shift and width. Two Gaussians are fit to the \HeII\ line to account for possible narrow and broad components of this line. The peak of the double-Gaussian fit is taken as the measured wavelength of \HeII, regardless of whether or not a narrow \HeII\ component is apparent in the spectrum. In most cases, only the broad \HeII\ component is well detected. However, if a strong, narrow \HeII\ component is present, the peak location of the double-Gaussian fit will be almost identical to that of the narrow component. 

\item[$\bullet$] \CIII\ is blended with \SiIII\ and \AlIII. We use an enforced symmetric profile for \CIII\ \citep{Shen_Liu_2012} to avoid confusion with the adjacent \SiIII\ line in the multi-Gaussian fit. This scheme is not perfect, but constrains the peak of the \CIII\ line complex well. We consider both the peak of the \CIII\ complex (including \CIII, \SiIII\ and \AlIII), denoted as \CIII$_a$,  and that of the decomposed \CIII\ line in the velocity measurements.  

\item[$\bullet$] Although a small subset of our spectra show evidence of a resolved \MgII\ doublet (at 2796.352\,\AA\ and 2803.531\,\AA\ corresponding to a velocity separation of $\sim 750\,\kms$) in emission, we fit \MgII\ as a single line in this work, using the single wavelength for \MgII\ listed in \citet{Vandenberk_etal_2001}. This is motivated by the practical use of \MgII\ as a redshift indicator, since most quasars do not show a resolved \MgII\ doublet given the typical line width (although some quasars with noticeable narrow \MgII\ emission could show resolved double \MgII\ peaks, the situation may be complicated by the potential associated absorption as well as spectral noise). For a similar reason we only fit a single Gaussian to \OII. 

\item[$\bullet$] For \hbeta, we only consider the broad component, as the narrow \hbeta\ component (subtracted from the model) is generally weaker than \OIII, and hence does not offer any advantage over \OIII\ when both lines are covered in the same spectral region. The broad \hbeta\ is, however, generally stronger than the narrow \OIII, and therefore is worth considering in cases where \OIII\ is too weak or noisy to measure. For example, in the case of extreme \FeII\ emitters, the \OIIIab\ lines are often swamped by \FeII\ emission at $\sim 4924$\,\AA\ and $5018$\,\AA\ \citep[e.g.,][]{Peterson_etal_1984}. We use \OIII$_c$ to denote the core component of \OIII, and \OIII$_a$ to denote the full \OIII\ profile. 

\end{enumerate}

We calculate the relative velocity shifts between two different lines using the model fits and the rest-frame wavelengths of the lines. By default we use the peak from the multi-Gaussian model as the fitted line center, but we consider alternative measures of the line center in \S\ref{sec:results} (e.g., centroid of the top percentile flux). The sign convention is such that negative velocity indicates blueshift, as measured in the observer's frame. 

We emphasize that the peak of the line is generally different from the centroid computed over the entire line profile. For narrow lines this distinction does not lead to a significant difference in the inferred line wavelengths. However, for broad lines the difference between the peak and the centroid calculated from the model fit can be substantial. For example, the centroid computed over the full broad \hbeta\ line has been found to display a net redshift rather than blueshift measured from the peak of broad \hbeta\ \citep[e.g.,][and references therein]{Tremaine_etal_2014}. This distinction should be kept in mind when applying the results from this work to other samples. 

To estimate the measurement uncertainties of the line centers, we adopted a Monte Carlo approach following earlier work \citep[e.g.,][]{Shen_etal_2008, Shen_etal_2011}: the original spectrum is randomly perturbed according to the formal flux density errors to create a mock spectrum; the same fitting approach is applied to the mock spectrum to derive the relevant spectral measurements; the measurement uncertainty of a particular quantity is then estimated as the semi-amplitude of the range enclosing the 16$^{\rm th}$ and 84$^{\rm th}$ percentiles of the distribution from 50 trials. Increasing the number of trials does not significantly change the estimated uncertainties. 

\section{Results}\label{sec:results}

\begin{figure*}
\centering
    \includegraphics[width=0.75\textwidth,angle=-90]{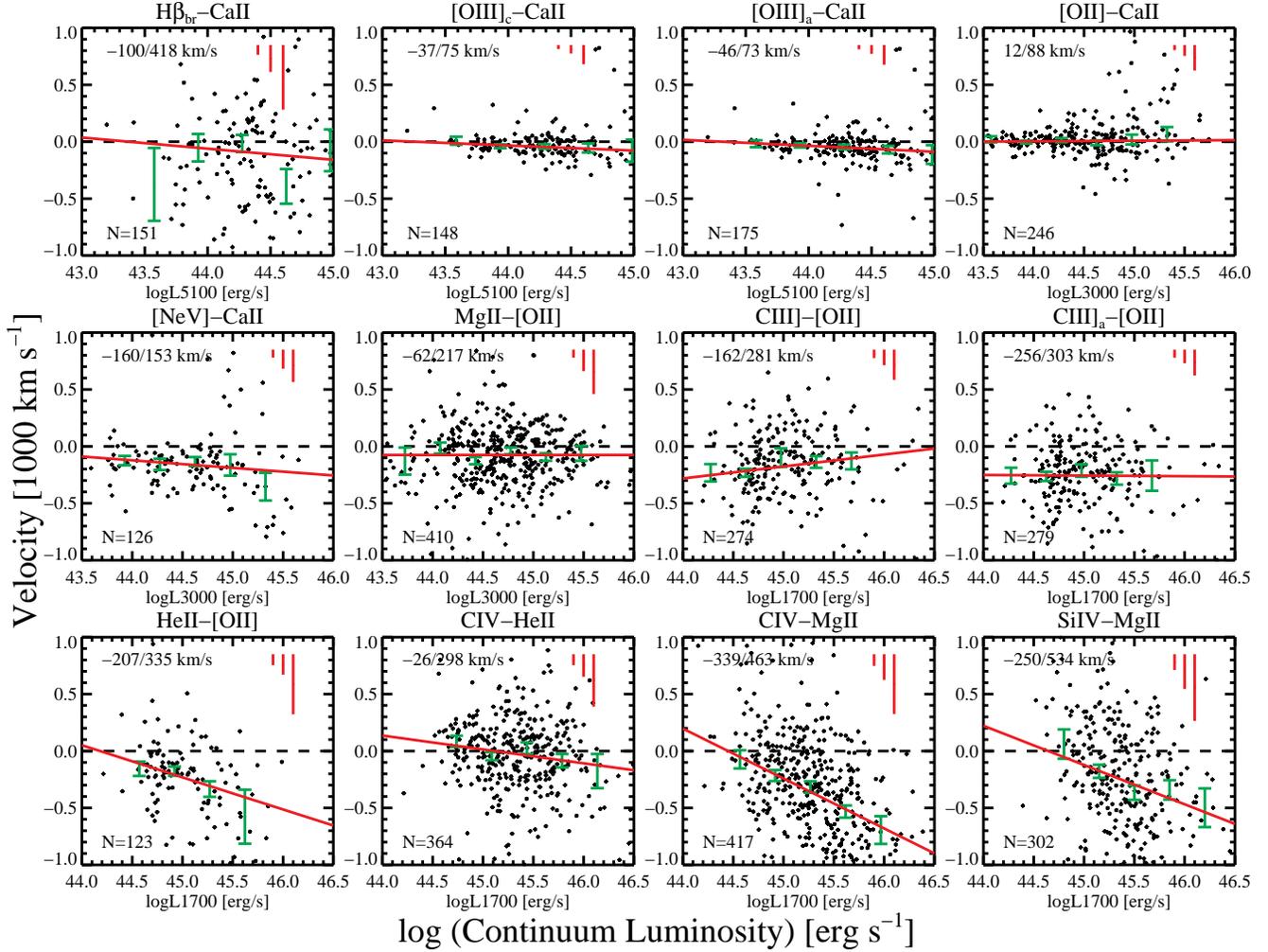}
    \caption{Velocity shift between two lines as a function of quasar continuum luminosity. Negative velocities indicate blueshift from systemic. Individual measurements are the black dots. The binned median value and uncertainty of the median in each luminosity bin are indicated in green bars. The red line is a linear fit to the average trend indicated by the binned points, with the best-fit parameters listed in Table \ref{tab:vel_lum}. The median and intrinsic scatter (see text) estimated from the full sample are marked in the top-left corner, and the sample size is indicated in the bottom-left corner. {The vertical red segments in the top-right corner are the 16th, 50th and 84th percentiles of the measurement uncertainties in the pairwise velocity shifts. Objects in our sample span a range in the measurement uncertainties in their line velocities (at $\sigma_{v}<500\,{\rm km\,s^{-1}}$), and these uncertainties generally increase as luminosity increases (e.g., higher-luminosity objects are on average at higher redshifts and fainter), leading to apparently more scatter at high luminosities in some panels (e.g., \OII-\CaII). However, as discussed in \S\ref{sec:line_shift}, the correlation analysis and binned linear fits are robust against a small fraction of noisy measurements. In particular, since the binned linear fit uses the median values in each bin, it is remarkably stable against apparent outliers. } }
    \label{fig:vel_shift}
\end{figure*}

\begin{table*}
\caption{Luminosity Trends}\label{tab:vel_lum}
\centering
\scalebox{1}{
\begin{tabular}{ccccccccccc}
\hline\hline
Line Pair & Luminosity & $\log L_{0}$ & $a$ & $b$ & $r$  & $r_{16\%}$ & $r_{84\%}$ &  $p$ & $p_{16\%}$ & $p_{84\%}$ \\
   & $\lambda L_\lambda$  &  ${\rm [erg\,s^{-1}]}$ & ${\rm [km\,s^{-1}]}$ & ${\rm [km\,s^{-1}]}$ \\
(1) & (2) & (3) & (4) & (5) & (6) & (7) & (8) & (9) & (10) & (11)\\
\hline
\hbeta$_{\rm br}$-\CaII  & $\log L_{5100}$ &  44 & $-61\pm75$ & $-98\pm174$ & $0.0028$ & $-0.078$ & 0.086 & 0.97 & 0.16 & 0.83 \\
\OIII$_{c}$-\CaII & $\log L_{5100}$ & 44 & $-33\pm12$ & $-45\pm39$ & $-0.11$ & $-0.20$  & $-0.022$ & $0.17$ & 0.015 &  0.66 \\
\OIII$_{a}$-\CaII & $\log L_{5100}$ & 44 & $-36\pm13$ &  $-53\pm35$ & $-0.15$ & $-0.23$ & $-0.073$ &  $0.044$ & $0.0023$ & 0.32 \\ 
\OII-\CaII & $\log L_{3000}$ & 44.5 & $6\pm12$  & $5\pm22$ & $0.067$ & $-0.0022$ & 0.13 & $0.31$ & 0.042 & 0.78 \\ 
\NeV-\CaII & $\log L_{3000}$ & 44.5 & $-156\pm25$  & $-66\pm61$ & $-0.23$ & $-0.32$ & $-0.12$ & $0.012$  & $3.4\times 10^{-4}$ & 0.18  \\ 
\MgII-\OII & $\log L_{3000}$ & 44.5 & $-76\pm18$ & $-0.1\pm40.2$ & $0.014$ & $-0.033$ & 0.062 & $0.78$ & 0.16 & 0.85\\ 
\CIII-\OII & $\log L_{1700}$ & 45 & $-176\pm 25$ & $105\pm 57$ & $0.11$ & 0.035 & 0.18 & $0.10$ & 0.0090 & 0.57 \\ 
\CIII$_a$-\OII & $\log L_{1700}$ & 45 &  $-256\pm 27$ & $-5\pm68$ & $-0.047$ & $-0.12$ & 0.018 & $0.48$ & 0.076 & 0.79 \\ 
\HeII-\OII & $\log L_{1700}$ & 45 & $-231\pm32$ & $-282\pm 123$ & $-0.40$ & $-0.49$ & $-0.31$ & $4.3\times 10^{-6}$ & $1.1\times 10^{-8}$ & $3.9\times 10^{-4}$ \\ 
\CIV-\HeII & $\log L_{1700}$ & 45 & $14\pm26$ & $-122\pm62$ & $-0.17$ & $-0.22$ & $-0.12$ & $1.4\times 10^{-3}$ & $1.9\times 10^{-5}$ & 0.024 \\ 
\CIV-\MgII & $\log L_{1700}$ & 45 & $-242\pm31$ & $-438\pm71$ & $-0.41$ & $-0.45$ & $-0.36$ & $5.7\times 10^{-18}$ & $1.1\times 10^{-21}$ & $1.5\times 10^{-14}$ \\ 
\SiIV-\MgII & $\log L_{1700}$ & 45 & $-123\pm55$ & $-345\pm104$ & $-0.35$ & $-0.41$ & $-0.31$ & $2.5\times 10^{-10}$ & $7.5\times 10^{-14}$ & $4.3\times 10^{-8}$ \\
\hline
\hline\\
\end{tabular}
}
\begin{tablenotes}
      \small
      \item NOTE. --- Best-fit parameters of Eqn.\ (\ref{eqn:L_trend}) on the binned data for different line pairs. For \OIII\ and \CIII\ the subscript ``$a$'' means the peak is measured from the entire line complex, and \OIII$_{c}$ refers to the core component of \OIII. Spearman's test results (rank correlation coefficient $r$ and null hypothesis probability $p$) on the individual data points are shown in Columns (6) and (9). {We also tabulate the 16th and 84th percentiles of the Spearman's $r$ and $p$ distributions from bootstrap resampling in Columns (7)-(8) and (10)-(11). }
\end{tablenotes}
\end{table*}

\subsection{Line Shifts}\label{sec:line_shift}

Fig.\ \ref{fig:vel_shift} presents the velocity shifts between lines as a function of continuum luminosity. We only consider lines that are detected at the $>3\sigma$ level. We further require that the line center is measured with an uncertainty of $<500\,\kms$ to reduce error-induced scatter in the measured velocity shifts. These particular line pairs were chosen to balance the need for good statistics (e.g., more objects in each comparison is better) and the usage of a relatively reliable systemic velocity indicator (e.g., \CaII, \OII, or \MgII) as redshift increases. In addition, some of the specific line pairs were chosen to serve as a reference for comparisons with other work (e.g., not all quasar samples can be used to reliably measure \CaII, {while \OII\ is relatively easier to measure in quasar spectra}). {In any case, we have confirmed that the results are mutually consistent when we swap the lines in these pairs. All the line velocity measurements used in this work are provided in an online FITS table. }

The velocity shift versus luminosity plots in Fig.\ \ref{fig:vel_shift} reveal different behaviors for different line species. Most line shifts show weak or no luminosity dependence, but \HeII, \CIV\ and \SiIV\ have strong luminosity dependence: their average blueshift relative to low-ionization lines such as \OII\ or \MgII\ increases when luminosity increases, consistent with previous findings \citep[e.g.,][]{Hewett_Wild_2010,Richards_etal_2011,Shen_etal_2011,Shen_Liu_2012}. When luminosity is fixed, there seems to be little redshift evolution in their average blueshifts, suggesting that luminosity is the main driver for the changes in line shifts \citep[see a similar argument for \OIII, e.g.,][]{Shen_2015}.

To quantify the average shift as a function of luminosity, we bin the data in luminosity (with a bin size of $\Delta\log L=0.35$) and present the median shift in each bin as green points in Fig.\ \ref{fig:vel_shift} with error bars estimated from the uncertainties in the median (standard deviation divided by $\sqrt{N}$, where $N$ is the number of objects in each bin).\footnote{The uncertainties in the bin median estimated this way are slightly larger than those derived from bootstrap resampling, possibly due to the outliers in each bin. We adopt the former as more conservative uncertainties in the bin median values and confirmed that the fitting results are consistent within 1$\sigma$ if we adopt the uncertainties in the median from bootstrap resampling in each bin.} {All data points are used in the binning, including a few objects that fall outside the plotting range in Fig.\ \ref{fig:vel_shift} for some line pairs.} We then fit a simple linear relation to the binned data points ({for robust bin-averaged values, bins with less than 5 objects are excluded in the fitting}):
\begin{equation}\label{eqn:L_trend}
v=a + b(\log L - \log L_0)\ ,
\end{equation}
where $\log L\equiv \log_{10}(\lambda L_{\lambda})$ is the specific monochromatic continuum luminosity (in logarithmic) at rest-frame wavelength $\lambda$ used for each line pair, and $\log L_{0}$ is a reference luminosity. In this framework a decade from the reference luminosity will lead to a change of $b\,\kms$ compared to the mean velocity shift at the reference luminosity. A larger $b$ value therefore indicates a stronger luminosity dependence in the velocity shift. The best-fit $a$ and $b$ values are listed in Table \ref{tab:vel_lum}. We also performed Spearman's test on the luminosity dependence of each line shift using all individual data points and the results are listed in Table \ref{tab:vel_lum}. The fitting results and the Spearman's tests confirm the visual impression from Fig.\ \ref{fig:vel_shift} that a luminosity dependence is highly statistically significant for the shifts of \HeII, \CIV, and \SiIV. The luminosity dependences of \HeII\ and \CIV\ shifts are consistent with those found in \citet{Denney_etal_2016b}. {For the other lines we do not detect statistically significant luminosity dependence with our sample, although some of them show evidence for a mild luminosity dependence such as \OIII\ and \NeV.}

{The reason for fitting the binned rather than individual data points is to reduce the impact of outliers.\footnote{Our tests in \S\ref{sec:sntest} suggest that the estimation of measurement errors in velocity shifts is reliable for most objects, but it is possible that the measurement errors may be underestimated for a few objects, causing them to be apparent outliers. However, the binned fits are insensitive to these few outliers. } Nevertheless, we also performed Ordinary-Least-Squares (OLS) ($Y|X$) fits and the Bayesian regression fits by \citet{Kelly_2007} on individual data points, and generally found consistent results as for the binned fits within 1-2$\sigma$. We further tested keeping only objects with better velocity measurements (uncertainties less than $50\,{\rm km\,s^{-1}}$), or rejecting objects with a pair-wise velocity offset larger than 3 times the scatter (estimated from the semi-amplitude of the range enclosing the 16th and 84th percentiles of the distribution) away from the median in the sample. In both cases we still found consistent results with the binned fits.} For all practical purposes, the best-fit linear relation on the binned data does a satisfactory job of removing the average luminosity trend, if any, for different line pairs. 

{On the other hand, the Spearman's tests on the individual data points should be treated with caution in the case of weak correlations (e.g., with a formal $p>10^{-3}$). It is possible that the Spearman's test is more sensitive to how the sample is constructed than the linear fit on binned data, particularly in terms of the $p$ value. To evaluate the significance of the Spearman's results in cases of weak correlations we perform the following tests. We first limit the sample to higher-quality measurements (e.g., velocity measurement uncertainties $<50\,{\rm km\,s^{-1}}$). We found that in this case the $p$ value can increase considerably in cases of weak correlations. For example, the Spearman's test yields $r=-0.21$ ($p=0.18$) for \NeV-\CaII, while the linear relation fit to the binned data yields $b=-64\pm 54$, fully consistent with the earlier result. Secondly, we note that the reported $p$ values in Table \ref{tab:vel_lum} typically have large uncertainties in the case of weak correlations. For example, for \NeV-\CaII, bootstrap resampling to estimate the uncertainty in the Spearman's test \citep[e.g.,][]{Curran_2014} yields a 16-84th percentile range of $p$ of [$3.4\times 10^{-4}$, 0.18] using the full sample. These tests suggest that the linear fit to the binned data is a more robust way to evaluate the statistical significance of a luminosity dependence of velocity shift in our sample than the Spearman's test. }  

{We note that there are some objects with apparent velocity shifts of several hundreds $\kms$ between the narrow lines and \CaII\ in Fig.\ \ref{fig:vel_shift} (particularly at high luminosities). We visually inspected these objects and found these are mostly noisy measurements due to low spectral quality, since our sample quality cut is $500\,\kms$ in the single-line velocity measurement errors. The online FITS table also reports the measurement uncertainties in line velocities to identify these objects and remove them from the analysis as desired. }

\begin{figure*}
\centering
    \includegraphics[width=0.75\textwidth,angle=-90]{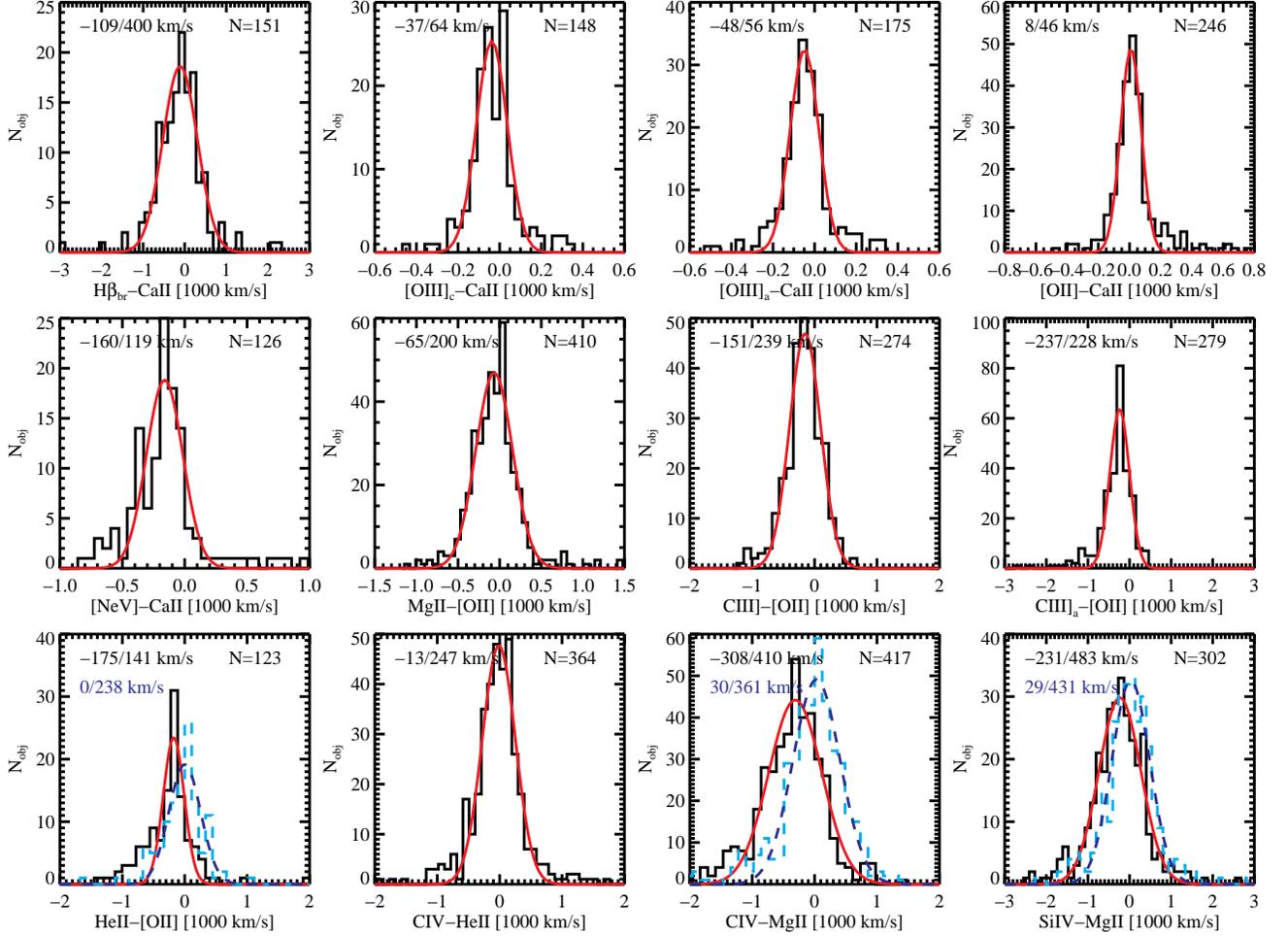}
    \caption{Distribution of velocity shifts (negative values indicating blueshift from systemic) for the line pairs shown in Fig.\ \ref{fig:vel_shift}. The red line is a Gaussian fit to the distribution, and the best-fit mean and intrinsic dispersion (best-fit Gaussian dispersion subtracting in quadrature the median measurement uncertainty in line shifts) are shown in the top-left corner. The total number of objects in each pair sample is shown in the top-right corner. A single Gaussian poorly fits the distribution of the \HeII-\OII\ velocity shift, and significantly underestimates the intrinsic scatter. For the velocity shifts of \HeII-\OII, \CIV-\MgII\ and \SiIV-\MgII, we also subtract the average luminosity dependence as shown in Fig.\ \ref{fig:vel_shift} with the best fit parameters listed in Table \ref{tab:vel_lum}. The corrected velocity shift distributions for the three line-pairs are indicated in the cyan dashed histogram, with the best-fit Gaussian models indicated by the blue dashed line. The best-fit Gaussian mean and measurement-error-corrected Gaussian dispersion are marked in blue in these panels. }
    \label{fig:vel_shift_hist}
\end{figure*}

Neglecting the (potential) luminosity dependence of the line shift, we display the histograms of the velocity shift for each pair of lines in Fig.\ \ref{fig:vel_shift_hist}, and fit a Gaussian to the distribution. The resulting mean and dispersion of the best-fit Gaussian are indicated in each panel in Fig.\ \ref{fig:vel_shift_hist}. We also measure the average shift and dispersion using the discrete points directly as shown in Fig.\ \ref{fig:vel_shift}: the median is used to denote the ``average'' shift, while the dispersion is estimated from the semi-amplitude of the range enclosing the 16$^{\rm th}$ and 84$^{\rm th}$ percentiles of the distribution, with the median measurement uncertainties in the velocity shift subtracted in quadrature. Both approaches yield consistent average shift and dispersion (intrinsic scatter) for the entire luminosity range probed, except for the \HeII-\OII\ shift, where the single-Gaussian fit failed to reproduce the highly non-Gaussian distribution. 


We list the mean and dispersion from the Gaussian fits in Table \ref{tab:vel_mean_scat}, where we also include additional fitting results for \HeII-\OII, \CIV-\MgII\ and \SiIV-\MgII\ after removal (subtraction) of the average luminosity trend. Other line shifts have negligible luminosity dependence, and thus do not require this refinement. We use these values as the ``average'' shift and intrinsic scatter for each line pair. Although a single Gaussian does not fit the velocity shift distribution perfectly (in particular in the wings of the distribution), the mean and dispersion from the Gaussian fits provide reasonable estimates for the average shift and the intrinsic scatter for the general population. 

Using the peak of the full line or the decomposed core component of \OIII\ yields similar mean shift and scatter. We have also tried the \OIII\ centroid computed from the line portion above half of the peak flux \citep[e.g.,][]{Hewett_Wild_2010}, and again found similar results. For \CIII, using the peak of the \CIII\ complex yields a more blueshifted velocity compared to using the peak of the decomposed \CIII. However, the blueshift of \CIII\ is luminosity-independent in both cases, and the scatter is also similar. Therefore in what follows we will consider the peaks measured from the full \OIII\ line and from the \CIII\ complex, as this approach does not suffer from potential incorrect or ambiguous decomposition of the line and therefore is more practical to use for redshift estimation. 

It is interesting to note that although \CIII\ and \CIV\ exhibit very different luminosity dependences of their velocity shifts, the \CIII\ FWHM (for the decomposed \CIII\ line) correlates with the \CIV\ FWHM well \citep[][]{Shen_Liu_2012}. This behavior suggests that pure kinematic processes (such as outflows) are probably not the only reason for \CIV\ blueshift, and radiative transfer effects may be responsible for the \CIV\ blueshift too. 

Tables \ref{tab:vel_lum}-\ref{tab:vel_mean_scat} provide empirical guidelines to infer the systemic redshift from specific lines and to quantify the intrinsic uncertainty in the redshift estimate, as described in \S\ref{sec:red}.

\begin{figure*}
\centering
    \includegraphics[width=0.75\textwidth,angle=-90]{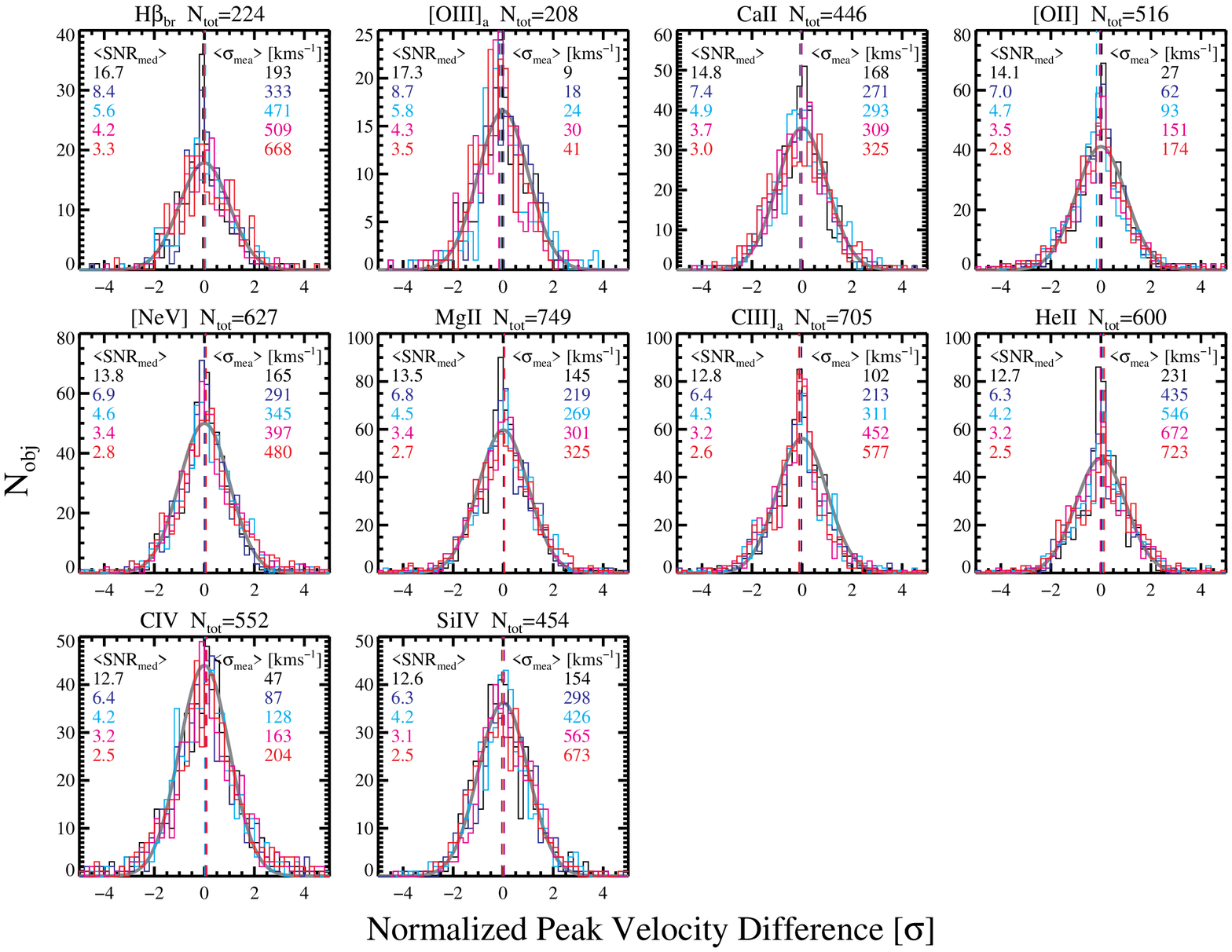}
    \caption{Distribution of the difference in the peak measurements between the original high SNR spectra and the degraded spectra, normalized by the measurement uncertainty (the quadrature sum of the measurement uncertainties in both measurements). Each panel shows results for a specific line, with the total number of objects listed at the top. Different colors represent different degradations of the original SNR by a factor of 2, 4, 6, 8 and 10. The sample median of the median SNR across the spectral range ($\bracket{{\rm SNR_{med}}}$) and the sample median measurement uncertainty ($\bracket{\sigma_{\rm mea}}$) in the degraded spectra are marked in each panel. The median of the distribution is indicated by the vertical dashed line. The gray solid line is a Gaussian with zero mean and unity dispersion, normalized to have the same area as the observed distribution. For all lines and all cases of SNR degradation, there is negligible offset in the distribution, indicating that the measurement of line peak is unbiased when the continuum SNR is decreased to as low as $\sim 3$ per SDSS pixel. In addition, the agreement between the observed distributions and the unity Gaussian suggests that our estimated measurement uncertainties from the Monte Carlo approach are reasonable. }
    \label{fig:vel_shift_sn}
\end{figure*}

\subsection{SNR Effects}\label{sec:sntest}

In practice we are often dealing with spectra with much lower SNR compared to the high-SNR coadded spectra considered here. One concern is that the measurement of the line peak may be subject to systematic biases at low SNR. To investigate the potential biases in line peak measurements as a function of SNR, we artificially degrade the spectral quality of our coadded spectra by adding noise at different levels, and perform the same fitting procedure to measure the line peaks \citep[e.g.,][]{Shen_etal_2011}. We adopt this approach instead of using the single-epoch spectra as low SNR representations of the high-SNR spectrum \citep[][]{Denney_etal_2016a} because many of our objects are at low redshifts (thus less $1+z$ time dilation) and hence line variability between the single-epoch and the final coadded spectra may introduce apparent differences in the measured line peaks. The test here specifically focuses on the effect of SNR, and not on the variability-induced velocity shifts (mostly on broad lines with short variability timescales) over the course of our monitoring program. 

We degrade the original spectral SNR by a factor of 2, 4, 6, 8 and 10 by randomly perturbing the original spectra with the corresponding error levels. We then re-measure the line peaks and compare to the original measurements made from the high SNR spectra. Fig.\ \ref{fig:vel_shift_sn} shows the distributions of differences between the measurements from the high SNR spectra and from the degraded spectra with the five error scales. The distributions are normalized by the measurement uncertainties, which are the quadrature sum of the measurement uncertainties from both the high SNR and the degraded SNR case (the latter naturally dominates the overall uncertainty). In order to retain the object-by-object diversity, we used all objects in our sample for this test, which span a range of SNR. We use the sample median of the degraded SNR (which itself is the median SNR across the full SDSS wavelength range) to quantify the quality of the spectra. 

Fig.\ \ref{fig:vel_shift_sn} suggests that there is no systematic bias in the measured line peaks when the spectral SNR is degraded to as low as $\sim 3$ per SDSS pixel, even though the measurement uncertainty in the line peak rises quickly when SNR decreases. In all cases the median of the distribution is $\lesssim 0.1\sigma$ and often much less than that, with no discernible trend with SNR. Considering the finite number of objects used for the distribution and the typical measurement uncertainties, the observed median offset between the high-SNR spectra and the degraded-SNR spectra is negligible in practice. 

This test also provides the typical measurement uncertainties in the line peak measurement as a function of SNR and for various lines, which are marked in each panel of Fig.\ \ref{fig:vel_shift_sn}. The trends with SNR and line species are well expected: at fixed SNR (mostly determined by the continuum), it is more difficult to measure the peak position precisely for weaker lines than for stronger lines, and the peak positions of the narrow lines can be measured to higher precision than those of the broad lines unless the narrow line is a weak line. 

Finally, Fig.\ \ref{fig:vel_shift_sn} indicates that our estimation of the measurement uncertainties in the line peaks from the Monte Carlo approach is reasonable, as the normalized difference distributions agree with a Gaussian distribution with unity dispersion reasonably well. In fact, in most cases this comparison suggests that our measurement uncertainties may have been slightly overestimated, as the observed distribution appears somewhat narrower than the Gaussian function. This result is probably caused by the fact that each Monte Carlo mock spectrum is the noise-free spectrum perturbed twice (i.e., the original spectrum already contains noise). 



\begin{table}
\caption{Mean and Scatter of Line Shifts}\label{tab:vel_mean_scat}
\centering
\scalebox{0.9}{
\begin{tabular}{ccccc}
\hline\hline
Line Pair & Mean & Intrinsic Scatter & Scatter w.r.t \CaII\ & median $\sigma_{\rm mea}$ \\
   & ${\rm [km\,s^{-1}]}$ & ${\rm [km\,s^{-1}]}$ & ${\rm [km\,s^{-1}]}$ & ${\rm [km\,s^{-1}]}$  \\
(1) & (2) & (3) & (4) & (5) \\
\hline
\hbeta$_{\rm br}$-\CaII  & $-109$  & 400  & ... &  119 \\
\OIII$_{c}$-\CaII  & $-37$  & 64  & ... & 37  \\
\OIII$_{a}$-\CaII & $-48$ & 56 & ... &  39 \\ 
\OII-\CaII & 8 &  46 & ...  & 48 \\ 
\NeV-\CaII  & $-160$ & 119 & ... & 84  \\ 
\MgII-\OII & $-65$ & 200 & 205 & 94 \\ 
\CIII-\OII & $-151$ & 239  & 243 & 69 \\ 
\CIII$_a$-\OII & $-237$ & 228 & 233 & 60 \\ 
\HeII-\OII & $-175$ & 238$^*$ & ... & 90\\
... & 0 & 238 & 242 & ... \\ 
\CIV-\HeII & $-13$ & 247 & ...  & 99  \\ 
\CIV-\MgII & $-308$ &  410 & ... & 113 \\
...  & 30 & 361 &  415 & ... \\ 
\SiIV-\MgII & $-231$ &  483 & ... & 153 \\
... & 29 & 431 & 477 & ... \\
\hline
\hline\\
\end{tabular}
}
\begin{tablenotes}
      \small
      \item NOTE. --- Mean velocity shift and intrinsic scatter for different line pairs. For \HeII-\OII, \CIV-\MgII\ and \SiIV-\MgII\ the second row lists the results after removing the average luminosity trend as shown in Fig.\ \ref{fig:vel_shift}. The last column lists the intrinsic scatter in the velocity shift w.r.t \CaII, if not already given in column (3). {The last column lists the median measurement uncertainty in the pair-wise velocity shifts.} \\
      $^*$The Gaussian fit to the uncorrected \HeII-\OII\ shift is poor and underestimates the true dispersion, therefore we use the intrinsic scatter inferred from the luminosity corrected distribution. 
\end{tablenotes}
\end{table}

\section{Empirical Recipes for Quasar Redshifts}\label{sec:red}

Following a similar approach as in \citet{Hewett_Wild_2010}, we now use the measured velocity shifts in \S\ref{sec:results} to construct a redshift ladder (a list of lines with decreasing accuracies in redshift estimation) to infer the systemic redshifts of quasars using various emission lines across the rest-frame optical-to-UV regime. 

Assume one has measured a ``redshift'' based on a particular line using the {\it peak} value measured from a model that accurately reproduces the data. A refined redshift can then be derived using the measured mean shifts in \S\ref{sec:results}. We emphasize that the model {\it must} provide a good fit to the overall line profile, otherwise there will be systematic differences when applying the mean shifts measured here. We generally recommend a multi-Gaussian function to fit the line given its flexibility, but other functional forms can be used as well. 

We can estimate how much difference there is in the inferred line peak velocity if the continuum and line decomposition is not performed on the spectrum. Assume the line is a Gaussian with a central wavelength $\lambda_0$ and a dispersion $\sigma_v$ (in velocity units), and has a rest-frame equivalent width of $W_0$ calculated using the continuum level at $\lambda_0$. Also assume the underlying continuum has a simple power-law form: $C_\lambda(\lambda)\propto (\lambda/\lambda_0)^{-\alpha}$. With a little algebra and to a very good approximation, the peak of the continnum$+$line flux will have a velocity offset from $\lambda_0$ as:
\begin{equation}
\frac{\Delta V}{c}\approx -\,\frac{\sqrt{2\pi}\alpha \lambda_0}{(1+z)W_0}\left(\frac{\sigma_v}{c}\right)^3\ ,
\end{equation} 
where $z$ is the redshift of the quasar and $c$ is the speed of light. For typical broad-line values (e.g., for broad \hbeta), $\alpha=1.5$, $\lambda_0=4862.68$\,\AA, $W_0=46$\,\AA, $z=1$, and $\sigma_v=2000\,{\rm km\,s^{-1}}$, we have $\Delta V\approx -18\,{\rm km\,s^{-1}}$. Similarly, for typical narrow-line values (e.g., for \OII) with $W_0=1.6$\,\AA, $\lambda=3728.48$\,\AA\ and $\sigma_v=250\,{\rm km\,s^{-1}}$ (other parameters held the same) we have $\Delta V\approx -0.8\,{\rm km\,s^{-1}}$. Such a velocity offset is negligible for essentially all practical uses. For similar reasons, for lines that lie above the \FeII\ complexes and for typical \FeII\ strengths, negligible velocity offsets will be induced in the line peak measurements if \FeII\ emission is not subtracted properly. However, we caution that if the line is weak while \FeII\ emission is strongly peaked near that emission line, decomposition of the \FeII\ emission will be necessary as it will affect the measurement of the peak of the line under consideration (e.g., for extreme optical \FeII\ emitters where \OIII\ is weak). 

We provide the following recommendations on redshift estimation for each line considered in this work. We assume that \CaII\ provides the most reliable systemic redshift, and other lines are directly or indirectly calibrated to \CaII\ using the redshift ladder from low to high redshifts, with as few intermediate lines as possible. When propagating the redshift uncertainties where one or more intermediate lines must be used to connect the line of interest to \CaII, we assume that the scatters in the pairwise velocity shifts are uncorrelated with each other so that the scatters are summed in quadrature to provide an estimate of the scatter relative to systemic. In practice this approach only affects \CIV\ and \SiIV\ below, where the intermediate line (\MgII) used displays a non-negligible (but still smaller) scatter with respect to systemic. Thus this approach provides a more conservative estimate of the redshift uncertainty for \CIV\ and \SiIV.


{\bf \OIII:} the peak of the full \OIII\ line provides a redshift estimate that is on average $-48\,\kms$ shifted from systemic, with an intrinsic uncertainty of 56$\,\kms$. Correcting for the marginal luminosity dependence of \OIII\ peak blueshift may lead to slight improvement in the average estimate at the high-luminosity end, following Eqn.\ (\ref{eqn:L_trend}) and Table \ref{tab:vel_lum}.  

{\bf \hbeta$_{\rm br}$:} the peak of broad \hbeta\ provides a redshift estimate that is on average $-109\,\kms$ shifted from systemic, with negligible luminosity dependence. The intrinsic uncertainty, however, is 400$\,\kms$, larger than those using narrow lines or some broad lines with typical line strength. 

{\bf \OII:} the peak of \OII\ from a single-Gaussian fit provides a redshift estimate that is on average shifted from systemic by only $+8\,\kms$, with no discernible luminosity dependence. The intrinsic uncertainty is only 46$\,\kms$. 

{\bf \NeV:} the peak of the full \NeV\,$\lambda$3426 line provides a redshift estimate that is on average shifted from systemic by $-160\,\kms$, with negligible luminosity dependence. The intrinsic uncertainty is 119$\,\kms$. 

{\bf \MgII:} the peak of full \MgII\ (broad$+$narrow components) provides a redshift estimate that is on average $-57\,\kms$ shifted from systemic, with negligible luminosity dependence. The intrinsic uncertainty is 205$\,\kms$ combining the scatter from \MgII-\OII\ and \OII-\CaII. 

{\bf \CIII:} the peak of the \CIII\ complex (including \CIII, \SiIII\ and \AlIII) provides a redshift estimate that is on average $-229\,\kms$ shifted from systemic, with negligible luminosity dependence. The intrinsic uncertainty is 233$\,\kms$ combining the scatter from \CIII-\OII\ and \OII-\CaII. 

{\bf \HeII:} the peak of the full \HeII\ line (broad$+$narrow components) provides a redshift estimate that is on average $-167\,\kms$ shifted from systemic. However, a luminosity dependence of the shift should be corrected using Eqn.\ (\ref{eqn:L_trend}) and the best-fit parameters listed in Table \ref{tab:vel_lum}. The corrected \HeII-based redshifts have an intrinsic uncertainty of 242$\,\kms$ combining the scatter from \HeII-\OII\ and (the smaller scatter from) \OII-\CaII.

{\bf \CIV:} After correcting for the luminosity dependence using Eqn.\ (\ref{eqn:L_trend}) and Table \ref{tab:vel_lum}, the peak of the full \CIV\ line provides a redshift estimate that is $-27\,\kms$ shifted from systemic, with an intrinsic uncertainty of 415$\,\kms$ combining the scatter from \CIV-\MgII\ and \MgII-\CaII.

{\bf \SiIV:} Using an effective rest wavelength of 1399.41\,\AA, and after correcting for the luminosity dependence using Eqn.\ (\ref{eqn:L_trend}) and Table \ref{tab:vel_lum}, the peak of the \SiIV/\OIV\ complex provides a redshift estimate that is on average $-28\,\kms$ shifted from systemic, with an intrinsic uncertainty of 477$\,\kms$ by combining the scatter in quadrature from \SiIV-\MgII\ and \MgII-\CaII. 


Once the residual constant offset is subtracted, the above various lines provide redshift estimates that are {\it on average} consistent with systemic, and an intrinsic uncertainty that depends on the line. Based on an increasing order of the intrinsic scatter, the emission-line list for redshift estimation in decreasing order of accuracy is: \OII, \OIII, \NeV, \MgII, \CIII, \HeII, broad \hbeta, \CIV, \SiIV. This list assumes that measurement uncertainties in the line centers are negligible.  

When multiple lines are available for redshift estimation, one also must consider the measurement uncertainties in each line and add the uncertainties in quadrature to the intrinsic scatter in velocity shifts. Weaker lines (such as \OII, \HeII) are presumably more difficult to measure precisely than stronger lines. The line redshift with the smallest overall uncertainties then provides the best estimate of the systemic redshift. It is also possible to average the redshift estimates from multiple lines, but since the uncertainties are likely correlated (from the velocity shift ladder) the resulting combined redshift uncertainty may underestimate the true uncertainty. 

\section{Physical Origins of Quasar Line Shifts}\label{sec:phy}

The observed average velocity shifts and scatter in different quasar emission lines contain information about the dynamical and radiative processes in these line-emitting regions. A detailed discussion of the physical origins of these velocity shifts is deferred to future work. Below we provide a brief and qualitative discussion on the observed line shifts. 

The strong luminosity dependence of the blueshifts of the high-ionization broad lines {\citep[e.g.,][]{Corbin_1990}} suggests that dynamical and/or radiative processes in the broad-line region associated with accretion are responsible for the observed blueshifts \citep[e.g.,][and references therein]{Richards_etal_2011}. {Models invoking outflows are often used to explain the observed \CIV\ blueshift \citep[e.g.,][]{Gaskell_1982, Chiang_Murray_1996, Richards_etal_2011}, but alternative interpretations exist \citep[e.g.,][]{Gaskell_Goosmann_2013, Gaskell_Goosmann_2016}.}

On the other hand, the kinematics of the narrow-line regions (such as outflows, rotation, mergers) coupled with anisotropic dust attenuation and/or orientation may account for most of the velocity shifts (both mean and scatter) observed in the high-ionization narrow lines such as \OIII\ and \NeV\ \citep[e.g.,][]{Veilleux_1991,Whittle_1992,Crenshaw_etal_2010}. 

A full understanding of these velocity shifts and their dependence on luminosity will require a simultaneous understanding of other line properties (such as line profile, equivalent width, line ratios, etc.) as functions of quasar parameters \citep[e.g.,][]{Baldwin_1977,Boroson_Green_1992,Sulentic_etal_2000,Dietrich_etal_2002,Baskin_Laor_2005,Dong_etal_2009,Richards_etal_2011,Zhang_etal_2011,Zhang_etal_2013,Stern_Laor_2012a,Stern_Laor_2012b,Stern_Laor_2013,Shen_Ho_2014,Luo_etal_2015,Shen_2015}. It is possible that secondary dependences of velocity shifts on other quasar properties can be used to further reduce the scatter in the observed shifts, which can be explored in future work with better samples. We also did not study subsamples of quasars with different multi-wavelength properties. For example, a small fraction ($\sim 10\%$) of quasars are radio loud, which have been shown to have smaller \CIV\ blueshifts on average {\citep[e.g.,][]{Steidel_Sargent_1991,Corbin_1992, Richards_etal_2011}}. The statistics of our sample does not provide better constraints on these subsamples of quasars compared with other studies that utilized much larger samples \citep[e.g.,][]{Richards_etal_2011}. 

While the low-ionization broad lines do not show a significant blueshift on average as opposed to the high-ionization broad lines, the scatter seen in their peak velocity shifts from systemic can be due to a variety of reasons, such as: (1) peculiar line profiles as predicted by certain broad-line region models \citep[e.g., disk emitters or disk winds,][]{Chen_etal_1989,Eracleous_Halpern_1994,Proga_etal_2000} {or off-axis illumination \citep[][]{Gaskell_2011}}; (2) line peak shifts due to broad-line region variability, or, in rare cases, a potential close black hole binary {\citep[e.g.,][]{Gaskell_1983,Eracleous_etal_2012,Shen_etal_2013,Ju_etal_2013,Liu_etal_2014}}; and (3) effects of reverberation with asymmetric transfer functions in velocity \citep[e.g.,][]{Barth_etal_2015}. These effects can lead to velocity shifts of up to hundreds of ${\rm km\,s^{-1}}$ or more on timescales of months to years \citep[e.g.,][]{Shen_etal_2013, Barth_etal_2015}. Similar processes may also contribute to the scatter in the velocity shifts of high-ionization broad lines. 



\section{Conclusions}\label{sec:sum}


We have studied the velocity shifts among the prominent narrow and broad emission lines in quasar spectra, and investigated the utility and uncertainties of using these individual lines as quasar systemic redshift indicators. This work is based on the spectroscopic quasar sample from the SDSS-RM project, which provided high SNR spectra to measure weak emission lines and stellar absorption lines that are generally difficult to measure in broad-line quasar spectra. The improvement of the current work over earlier studies is two-fold: 1) we were able to measure robust stellar \CaII\ absorption redshifts for a large sample of quasars that extends to $z\sim 1.5$, whereas earlier samples based on SDSS quasars were only able to measure \CaII\ in quasars at substantially lower redshifts and luminosities; 2) our sample spans a large dynamic range in luminosity ($\sim 2$ dex), allowing an investigation of the luminosity dependence of these line shifts to the high-luminosity end more appropriate for high-redshift quasars. 

The findings from this work on the velocity shifts in quasar emission lines and their implications for redshift estimation are summarized as follows:

\begin{enumerate}

\item[1.] High-ionization forbidden lines are generally more blueshifted than low-ionization forbidden lines (Fig.\ \ref{fig:vel_shift}). In particular, \NeV\ is more blueshifted than \OIII, and \OIII\ is more blueshifted than \OII. There is some evidence that the \OIII\ peak blueshift mildly depends on quasar luminosity, with the average blueshift potentially reaching $\sim 100\,\kms$ at $\log (L_{5100}/{\rm erg\,s^{-1}})=45$, although our sampling at such high luminosities is sparse. This is roughly consistent with the claim by \citet{Marziani_etal_2015} that the fraction of \OIII\ blue outliers (e.g., \OIII\ peak blueshifts) increases with luminosity.

\item[2.] Similarly, high-ionization permitted lines are generally more blueshifted than low-ionization permitted lines. The blueshifts of \HeII, \CIV\ and \SiIV\ are strongly luminosity-dependent, with higher luminosity quasars showing larger blueshifts in these lines. On the other hand, the shift of \MgII\ is mild and luminosity-independent. \CIII\ shows a significant blueshift that does not depend on luminosity. However, if a single Gaussian were fit to the \CIII\ complex, the relative strength between \SiIII\ and \CIII\ as a function of luminosity would lead to a luminosity dependence of the apparent \CIII\ blueshift \citep[e.g.,][]{Richards_etal_2011}.

\item[3.] The measured velocity shifts among different quasar emission lines were used to derive empirical recipes for systemic redshift estimation based on various lines from low to high redshifts. These empirical recipes account for the average velocity shifts of each line with respect to systemic, and therefore provide unbiased redshift estimation on average. More importantly, we quantified the typical uncertainties of these redshift estimates based on various lines, using the observed intrinsic scatter in the velocity shifts. 

\item[4.] Consistent with recent work \citep[e.g.,][]{Hewett_Wild_2010,Shen_etal_2011,Shen_2015}, we found that \MgII\ provides the best broad-line redshift estimates, with an intrinsic uncertainty of $\sim 200\,\kms$. We also found that \CIII\ (after correcting for the average velocity shift) provides only slightly more uncertain redshift estimates. {Based on a different approach, Allen \& Hewett (in preparation) used mean field
independent component analysis (MFICA) \citep[see][for an application
to astronomical spectra]{Allen_etal_2013} to reconstruct the \MgII\ and \CIII\ regions, and derived \MgII- and \CIII-based redshifts with similar intrinsic uncertainties of $\sim 230\,\kms$, with no detectable luminosity dependence in the mean velocity shifts. } 

\item[5.] For \HeII, \CIV\ and \SiIV, after removing the average luminosity trends, these three permitted lines provide unbiased redshift estimation on average, but with typical redshift uncertainties of several hundreds of $\kms$. Although the intrinsic scatter for \HeII\ is only slightly larger than that for \CIII, it is generally not as favorable as \CIII\ for redshift estimation, both due to the luminosity dependence and the fact that \HeII\ is much weaker than \CIII. 

\end{enumerate}

\acknowledgements


We thank the referee for useful comments, and Paul Hewett for helpful discussions. YS acknowledges support from an Alfred P. Sloan Research Fellowship. WNB and CJG acknowledge support from NSF grant AST-1517113. KDD is supported by an NSF AAPF fellowship awarded under NSF grant AST-1302093. BMP is grateful for support by the NSF through grant AST-1008882 to the Ohio State University. 

Funding for SDSS-III has been provided by the Alfred P. Sloan Foundation, the
Participating Institutions, the National Science Foundation, and the U.S.
Department of Energy Office of Science. The SDSS-III web site is
http://www.sdss3.org/.

SDSS-III is managed by the Astrophysical Research Consortium for the
Participating Institutions of the SDSS-III Collaboration including the
University of Arizona, the Brazilian Participation Group, Brookhaven National
Laboratory, University of Cambridge, Carnegie Mellon University, University
of Florida, the French Participation Group, the German Participation Group,
Harvard University, the Instituto de Astrofisica de Canarias, the Michigan
State/Notre Dame/JINA Participation Group, Johns Hopkins University, Lawrence
Berkeley National Laboratory, Max Planck Institute for Astrophysics, Max
Planck Institute for Extraterrestrial Physics, New Mexico State University,
New York University, Ohio State University, Pennsylvania State University,
University of Portsmouth, Princeton University, the Spanish Participation
Group, University of Tokyo, University of Utah, Vanderbilt University,
University of Virginia, University of Washington, and Yale University.


\end{document}